\documentclass[twocolumn,a4paper,prl]{revtex4-1}

\usepackage{graphicx}
\usepackage{graphics}
\usepackage{amsmath}
\usepackage{amssymb}
\usepackage{amsfonts}
\usepackage{dsfont}
\usepackage{braket}
\usepackage{color}
\usepackage{natbib}

\date{\today}

\begin{document}

\title{Particles, holes and solitons: a matrix product state approach}
\author{Damian Draxler$^1$, Jutho Haegeman$^{2}$, Tobias J.~Osborne$^3$, Vid Stojevic$^{12}$, Laurens Vanderstraeten$^2$, Frank Verstraete$^{12}$}
\affiliation{$^1$ Vienna Center for Quantum Science, Universit\"at Wien, Boltzmanngasse 5, A-1090 Wien, Austria\\
$^2$ Ghent University, Department of Physics and Astronomy, Krijgslaan 281- S9, B-9000 Ghent, Belgium\\
$^3$ Leibniz Universit\"{a}t Hannover, Institute of Theoretical Physics and Riemann center for geometry and physics, Appelstrasse 2, D-30167 Hannover, Germany}

\begin{abstract}
We introduce a variational method for calculating dispersion relations of translation invariant (1+1)-dimensional quantum field theories. The method is based on continuous matrix product states and can be implemented efficiently. We study the critical Lieb-Liniger model as a benchmark and excelent agreement with the exact solution is found. Additionally, we observe solitonic signatures of Lieb's Type  \MakeUppercase{\romannumeral 2} excitation. In addition, a non-integrable model is introduced where a $U(1)$-symmetry breaking term is added to the Lieb-Liniger Hamiltonian. For this model we find evidence of a non-trivial bound-state excitation in the dispersion relation. 
\end{abstract}

\maketitle

The last decades have witnessed an explosion in the experimental realization of strongly correlated one-dimensional quantum systems \citep{Giamarchi:2003fk}. Often a discretised description in terms of a lattice Hamiltonian is possible, which can then be studied using White's density matrix renormalization group \citep{PhysRevLett.69.2863}. The underlying variational ansatz, the set of matrix product states (MPS) \citep{PhysRevLett.59.799,Fannes:1992uq,PhysRevLett.75.3537,doi:10.1080/14789940801912366,1751-8121-42-50-504004}, explains both the success of this method and has been used to develop generalizations beyond the setting of ground states, \textit{e.g.}\ to the study of time evolution \citep{PhysRevLett.93.040502,PhysRevLett.107.070601}, dissipative dynamics \citep{PhysRevLett.93.207204,PhysRevLett.93.207205} and dispersion relations \citep{PhysRevB.85.035130,PhysRevB.85.100408}.

The use of elongated optical or magnetic atom traps has opened the possibility of creating one-dimensional quantum gases in the lab  \citep{PhysRevLett.87.130402,PhysRevLett.91.250402,Kinoshita20082004,PhysRevLett.95.190406,PhysRevLett.96.130403,PhysRevLett.105.265302}. It is natural to study these systems directly using quantum fields, without resorting to a lattice discretisation. A continuum limit of the class of matrix product states, known as \textit{continuous matrix product states} (cMPS)  \citep{PhysRevLett.104.190405,PhysRevLett.105.260401,2012arXiv1211.3935H}, was recently developed and has demonstrated to be able to provide an efficient description of the ground-state properties of the Lieb-Liniger (LL) model \citep{PhysRev.130.1605}.

Apart from ground-state properties, there has also been experimental interest in localized excitations in these systems \citep{PhysRevLett.83.5198,2008NatPh...4..496B,PhysRevLett.101.120406,PhysRevLett.101.130401}. While Lieb determined the spectrum of excitations for the LL model \citep{PhysRev.130.1616}, a systematic method for studying excitations of non-integrable quantum fields is still lacking. In this Letter we fill the gap by extending the recently introduced ansatz for excitations of translation invariant spin chains in the thermodynamic limit \citep{PhysRevB.85.100408} to the setting of cMPS. This yields a new variational ansatz for  elementary excitations of translation invariant quantum fields that allows us to simulate dispersion relations for integrable and non-integrable models alike. The corresponding variational states are faithful eigenstates and have therefore an infinite lifetime, out of which we can construct localized wavepackets by taking linear combinations. With this method, we can reconstruct the spectrum of the LL Hamiltonian and illustrate the solitonic effects in Lieb's Type \MakeUppercase{\romannumeral 2} excitation ---first observed in Refs.~\onlinecite{1976TMP....28..615K,Ishikawa} in the weak-interaction limit--- for arbitrary interaction strength. We can equally well construct the spectrum for non-integrable models, which we illustrate by adding a pairing term to the LL Hamiltonian which opens a gap. For a certain parameter regime, our method provides strong evidence for the existence of a non-trivial bound state. 

A cMPS for a translation invariant infinite system with open boundary conditions is defined as \cite{PhysRevLett.104.190405}
\begin{equation}
|\Psi(Q, R)\rangle = v_{L}^{\dagger} \left( \mathcal{P}\mathrm{e}^{\int_{-\infty}^{\infty} \mathrm{d}x\, [ Q\otimes\mathds{1} + R\otimes\hat{\psi}^{\dagger}(x) ]} \right) v_{R} |\Omega\rangle \;, \nonumber
\end{equation}
where $Q, R \in \mathds{C}^{D\times D}$, $ v_{L}$ and $v_{R} $ are $D$-dimensional boundary vectors acting on an ancillary system, $|\Omega\rangle$ is the Fock vacuum and $\mathcal{P}$ the path-ordering operator. For bosonic systems the field operators satisfy the commutation relation $[\hat{\psi}(x),\hat{\psi}^{\dagger}(y)] = \delta (x-y)$.  For a generic normalizable cMPS, all eigenvalues of the transfer matrix $T=Q\otimes\mathds{1}+\mathds{1}\otimes\bar{Q}+R\otimes\bar{R}$ have non-positive real part and there is a non-degenerate zero eigenvalue. The corresponding left and right zero eigenvectors, $\langle l|T=0$, $T|r\rangle=0$ may be reshaped to give positive, hermitian matrices $l$ and $r$ which have full rank and are normalized so that $\langle l|r\rangle = \textnormal{Tr}(lr) = 1$. Since the boundary vectors have no variational importance they are chosen so that the state has norm $1$, i.e.   $\langle\Psi(\bar{Q},\bar{R})|\Psi(Q,R)\rangle = (v_{L}^{\dagger}\otimes v_{L}^{\top})|r\rangle\langle l|\left(v_{R}\otimes\bar{v}_{R}\right) = (v_{L}^{\dagger} r v_{L})(v_{R}^{\dagger} l v_{R}) = 1$.

Suppose we have approximated the ground state of a system as a cMPS parametrized by the matrices $(Q,R)$. An ansatz for a particle-like eigenstate or excitation is obtained by locally replacing the matrices $Q$ and $R$ by $V$ and $W$ --- which has the effect of perturbing the ground state in a spatial region of the size of the correlation length --- and then building a state with definite momentum via a plane wave superposition:
 \begin{equation}
 \begin{split}
 |\Phi_{p}&(V,W)\rangle := \int_{-\infty}^{\infty} \mathrm{d}x \,\mathrm{e}^{\mathrm{i}px} v_{L}^{\dagger}\hat{U}_{1}(-\infty,x) \\
  &\times \Big( V\otimes\mathds{1} + W\otimes\hat{\psi}^{\dagger}(x) \Big) \hat{U}_{2}(x,\infty)v_{R}|\Omega\rangle \;, \nonumber
 \end{split}
 \end{equation}
where $\hat{U}_{a}(x,y)=\mathcal{P}\exp(\int_{x}^{y}\mathrm{d}z\,(Q_{a}\otimes\mathds{1}+R_{a}\otimes\hat{\psi}^{\dagger}(z)))$ for $a=1,2$. The crucial feature here is a single strictly local disturbance described by the matrices $V,W\in\mathds{C}^{D\times D}$, which is nevertheless able to influence the state up to a distance determined by the bond dimension and appears to efficiently capture single-particle excitations, as we illustrate below. The states $\ket{\Phi_{p}(V,W)}$ are momentum eigenstates and obey a $\delta$-orthogonality. Note that they depend linearly on the variational parameters $V$ and $W$, so they span a linear subspace of Hilbert space. Asymptotically the states $ \ket{\Phi_{p}(V,W)}$ look like $\ket{\Psi(Q_1,R_1)}$ at $x=-\infty$ and $\ket{\Psi(Q_2,R_2)}$ at $x=+\infty$, which are supposed to be equally good but potentially different ground states (in the case of symmetry breaking). Our ansatz thus includes the possibility of capturing \textit{topologically non-trivial} excitations, henceforth referred to as topological excitations for the sake of brevity. For $\ket{\Psi(Q_1,R_1)}=\ket{\Psi(Q_2,R_2)}$, the ansatz describes (topologically) trivial excitations and can be interpreted as a state in the momentum sector $p$ of the tangent space obtained by an infinitesimal position dependent variation of $\ket{\Psi(Q,R)}$ \cite{2012arXiv1211.3935H}.

To compute excitations we apply the \textit{Rayleigh-Ritz} method, which results in a generalized eigenvalue problem for an effective Hamiltonian. In our case the generalized eigenvalue equation is given by
 \begin{equation}
H_{p}\begin{bmatrix} V \\ W \end{bmatrix} = E N_{p}\begin{bmatrix} V \\ W \end{bmatrix} \;,\nonumber
\end{equation}
with $[\begin{smallmatrix}V \\ W\end{smallmatrix}]$ being the $2D^{2}$-dimensional vector corresponding to $V$ and $W$, $E$ is the energy, $H_{p}$ the effective Hamiltonian and $N_{p}$ the effective norm matrix. Both $H_{p}$ and $N_{p}$ are $2D^{2}\times 2D^{2}$-dimensional matrices  defined by
\begin{small}
\begin{align}
\langle\Phi_{p}(\overline{V},\overline{W})|\hat{H} - E_{0}|\Phi_{p^{\prime}}(V,W)\rangle = 2\pi\delta(p-p^{\prime})\begin{bmatrix} V^{\dagger} W^{\dagger}\end{bmatrix} H_{p}  \begin{bmatrix} V \\ W \end{bmatrix} \nonumber \\
\langle\Phi_{p}(\overline{V},\overline{W})|\Phi_{p^{\prime}}(V,W)\rangle = 2\pi\delta(p-p^{\prime})\begin{bmatrix} V^{\dagger} W^{\dagger}\end{bmatrix} N_{p}  \begin{bmatrix} V \\ W \end{bmatrix} \;, \nonumber
\end{align}
\end{small}
where $E_0$ is the ground state energy obtained with the ground state approximation $\ket{\Psi(Q,R)}$.

Note that $H_{p}$ and $N_{p}$ have zero eigenvalues corresponding to a redundancy in the representation of the states $\ket{\Phi_p(V,W)}$, which can be traced back to the gauge invariance of the states $\ket{\Psi(Q,R)}$ under the transformation $Q\leftarrow g^{-1} Q g$ and $R\leftarrow g^{-1} R g$ \cite{2012arXiv1211.3935H}. One finds that  for all $X\in\mathds{C}^{D\times D}$, $|\Phi_p([X,Q] + \mathrm{i}pX,[X,R])\rangle = 0$. Hence,  $|\Phi_p(V^{\prime},W^{\prime})\rangle=|\Phi_p(V,W)\rangle$ if $V^{\prime}=V+[X,Q]+\mathrm{i}pX $ and $W^{\prime}=W+[X,R]$. If, for $p=0$, one also restricts to states orthogonal to the ground state by imposing $\langle l| \left( V\otimes\mathds{1} + W\otimes\bar{R} \right)|r\rangle=0$, there are $D^2$ redundant degrees of freedom for every momentum $p$. These can be eliminated by constraining $V$ and $W$ to satisfy a `gauge-fixing' condition such as
\begin{equation}
\langle l| \left( V\otimes\mathds{1} + W\otimes\bar{R} \right) = 0 \;,
\label{eq:gaugefixing}
\end{equation}
with $\langle l|$ the left eigenvector of $T_{11}$, the transfer matrix corresponding to $\ket{\Psi(Q_1,R_1)}$. It can be shown that this choice of gauge reduces the effective norm matrix to the identity (Ref.~\onlinecite{2012arXiv1211.3935H} and Supplementary Material), so that the Rayleigh-Ritz problem becomes an ordinary eigenvalue problem. The explicit calculation of the effective Hamiltonian $H_{p}$ in this gauge is more involved and is derived in full detail in the Supplementary Material. The lowest eigenvalues of $H_{p}$ can then be obtained with a computational time scaling as $\mathcal{O}(D^3)$ using a sparse eigensolver exploiting the tensor product structure of the effective Hamiltonian.

\begin{figure}
\centering
\includegraphics[width=85mm,height=44mm]{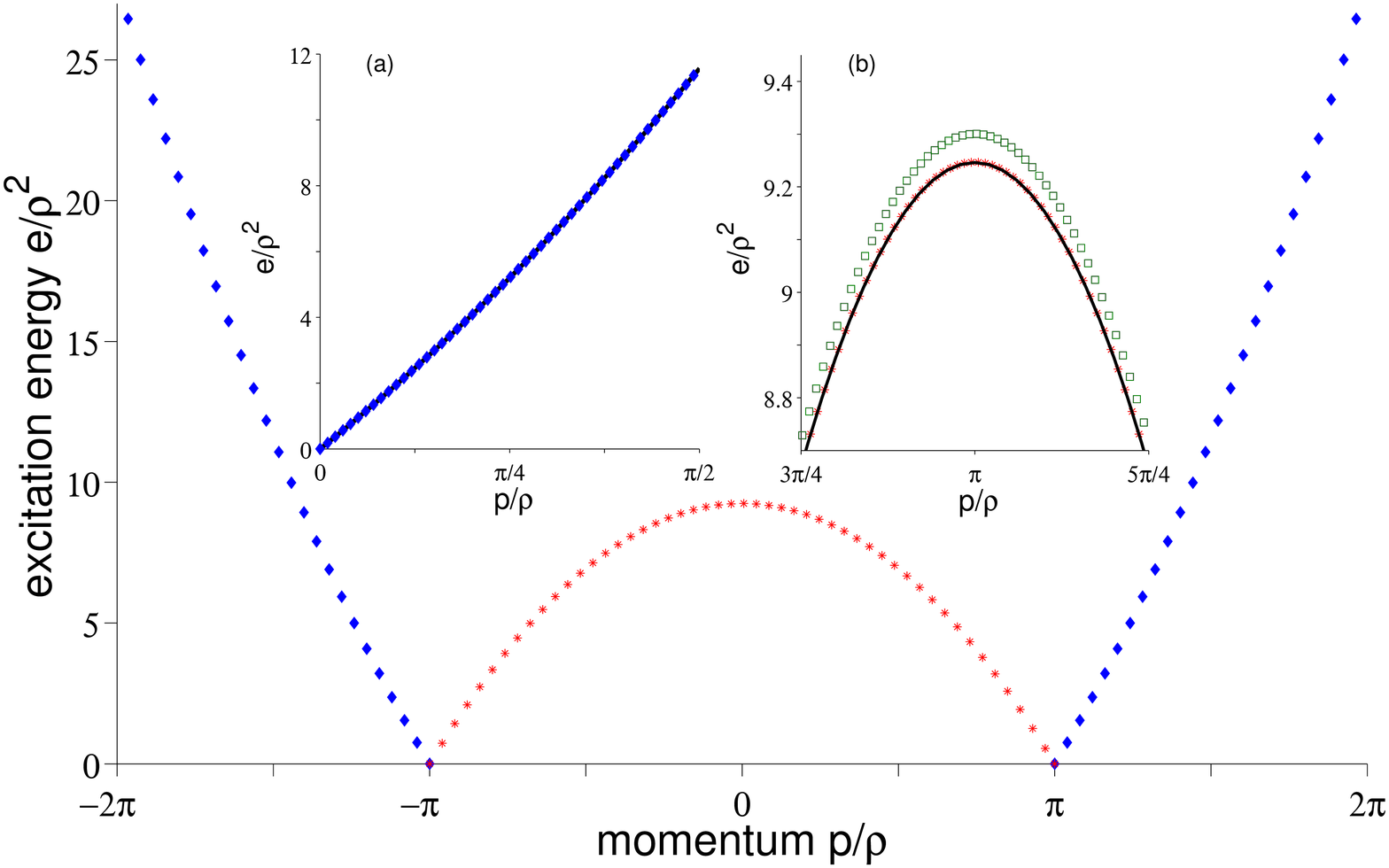}
\caption{\footnotesize{Elementary topological excitation branches for $\gamma=c/\rho\approx 60.16$ at $D=64$ with $\theta=\pi$. Blue diamonds are \textit{particle} excitations and red stars \textit{hole} excitations. (a) Dispersion relation of Lieb's Type  \MakeUppercase{\romannumeral 1} excitation obtained  from the topological ansatz (blue diamonds) and from the Bethe ansatz solution (black line). (b) Dispersion relation of Lieb's Type \MakeUppercase{\romannumeral 2} excitation obtained from the topological ansatz (red stars), the trivial ansatz (green squares) and from the Bethe ansatz solution (black line).}}
\label{fig:LL2}
\end{figure}
\begin{figure}
\centering
\includegraphics[width=85mm,height=45mm]{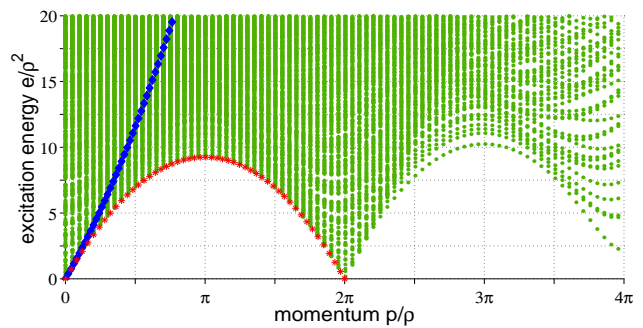}
\caption{\footnotesize{Dispersion relation for the Lieb-Liniger model with $D=64$ for $\gamma =c/\rho\approx  60.16$. Bulk excitations (dots) are trivial excitations. Blue diamonds (Type \MakeUppercase{\romannumeral 1}) and red stars (Type \MakeUppercase{\romannumeral 2}) are obtained by combining momenta and energy of two topological  excitations (hole and particle) of Fig.~\ref{fig:LL2}.}}
\label{fig:LL}
\end{figure}

We now discuss the results obtained for the LL model. The LL Hamiltonian is given by
\begin{equation}
\hat{H}_{\mathrm{LL}}=\int_{-\infty}^{+\infty} \left[\frac{\mathrm{d} \hat{\psi}^{\dagger}}{\mathrm{d} x}\frac{\mathrm{d} \hat{\psi}}{\mathrm{d} x}-\mu \hat{\psi}^{\dagger}\hat{\psi}+c \hat{\psi}^{\dagger}\hat{\psi}^{\dagger}\hat{\psi}\hat{\psi}\right]\mathrm{d}x,
\end{equation}
with a repulsive interaction strength $c>0$ and where the argument of the field operators has been omitted for the sake of brevity. Since a variational approach targets the lowest energy state, this Hamiltonian was formulated in the grand-canonical ensemble (with chemical potential $\mu>0$). The Hamiltonian is gapless and only depends on a single parameter $\gamma = c/\rho$, with $\rho$ the ground state particle density which is set by $\mu$. The cMPS ansatz generalizes a coherent state ansatz and breaks the U(1)-symmetry of the model, whereas the exact Bethe-ansatz ground state does not and has a fixed total number of particles. This follows from the fact that it is often energetically beneficial to break the symmetry in the presence of a constraint on the total amount of entanglement. Hence, the order parameter $\langle\Psi(\bar{Q},\bar{R})|\hat{\psi}|\Psi(Q,R)\rangle \neq 0$  (and in fact slowly converges to $0$ for increasing $D$). For any $\theta \in[0,2\pi)$, $|\Psi(Q,e^{i\theta}R)\rangle$ is again a valid ground state so that we can also consider topological excitations interpolating between two different ground states characterized by a different order parameter. They can also be understood as momentum superpositions of a local perturbation at position $x$ which has a half-infinite string $\hat{S}(x)=\exp(\mathrm{i}\theta \int_{-\infty}^{x} \hat{\psi}^{\dagger}(
z)\hat{\psi}(z)\,\mathrm{d}z)$ attached to it. Even for the exact solution with fixed particle number $N$, the elementary particle ($N+1$) and hole ($N-1$) excitations have a topological nature and need to be studied using antiperiodic boundary conditions \citep{Korepin:1997kx}. For $\theta=\pi$, the string $\hat{S}(x)$ has exactly the effect of flipping the sign of the field operators at $-\infty$ ($\hat{S}(x)^{\dagger} \hat{\psi}(-\infty)\hat{S}(x)=-\hat{\psi}(-\infty)$) and Fig.~\ref{fig:LL2} does indeed confirm that the elementary excitations are perfectly captured by using the ansatz for topological excitations with the choice $Q_{1}=Q_2=Q$, $R_{1}=-R_{2}=R$. The dispersion relation is centered around $0$ and the \textit{hole} branch (red stars) is obtained as the lowest excitation energy with momentum between $-\pi\rho$ and $+\pi\rho$, with $\rho$ the particle density. The \textit{particle} branch (blue diamonds) shows the eigenvalues of the eigenvectors that have the largest overlap with the state $\int_{-\infty}^{\infty}\mathrm{d}x\,\mathrm{e}^{\mathrm{i} p x} \hat{\psi}^{\dagger}(x) e^{\mathrm{i}\pi\int_{-\infty}^{x}\hat{\psi}^{\dagger}(z)\hat{\psi}(z)\,\mathrm{d}z}|\Psi(Q,R)\rangle$. All parameters plotted in the figures are normalized such that they are dimensionless ($p/\rho, e/\rho^2$), as in Ref. \onlinecite{PhysRev.130.1605}. 

Lieb determined the spectrum with fixed particle number (\textit{i.e.} topologically trivial) in first quantization \citep{PhysRev.130.1616} and isolated two types of excitations, which he labeled Type \MakeUppercase{\romannumeral 1} and Type \MakeUppercase{\romannumeral 2} excitations. Either can be used to construct the full spectrum of excitations with equal particle number. Fig.~\ref{fig:LL} shows the eigenvalues of the effective Hamiltonian $H_p$ as a function of the momentum $p$ obtained using the trivial ansatz ($Q_1=Q_2$ and $R_1=R_2$). However, it is well-known \citep{Korepin:1997kx} that the Type \MakeUppercase{\romannumeral 1} excitations should be understood as one \textit{hole} at momentum $-\pi\rho$ plus one \textit{particle} with momentum $p \geq \pi\rho$, whereas the Type \MakeUppercase{\romannumeral 2} excitations at momentum $p$ are obtained by combining one \textit{particle} with momentum $\pi\rho$ plus one \textit{hole} with momentum $-\pi\rho\leq p\leq\pi\rho$. By combining momentum and variational energies of our topological excitations according to this recipe, we can accurately reproduce ($5$ digits of precision) the Bethe ansatz dispersion relations of Lieb's Type I [blue diamonds in Fig.~\ref{fig:LL} and \ref{fig:LL2}(a)] and II excitations [red stars in Fig.~\ref{fig:LL} and \ref{fig:LL2}(b)]. If we would have used the trivial excitation energies directly [also sketched in Fig.~\ref{fig:LL2}(b)], a worse precision would have been obtained, as this single particle ansatz cannot accurately represent the two unbound constituents of Lieb's Type I and Type II excitations and tries to confine them into a small spatial region. Since the repulsive Lieb-Liniger model does not have any bound states \citep{Korepin:1997kx}, all trivial excitations have a similar structure consisting of an even sum of topological hole and particle excitations.

Without knowing the exact result, this information could be inferred from looking at the convergence of the variational energies as a function of increasing $D$. There are two competing effects within our variational strategy. Firstly, by modifying $Q$ and $R$ locally, we assume that the excitation is confined in a spatial region, the width of which is set by the bond dimension. This is the variational approximation and it produces a positive variational error. Secondly, we subtract from the Hamiltonian a variational estimate of the ground state energy which is too large. This second effect results in a negative error and is dominant for truly elementary excitations for which the assumed locality is valid and the first error is negligible. If an excited state cannot be approximated locally, \textit{e.g.}\ an excitation that is composed of several unbound elementary excitations, then the first effect will likely be dominant as the ansatz confines these different excitations in a finite spatial region. Indeed, for increasing $D$, the particle and hole energies in Fig.~\ref{fig:LL2} are increasing (which is true for all $\gamma\gtrsim3$) because a smaller variational estimate for the ground state energy is subtracted. In contrast, the energies of trivial excitations in Fig.~\ref{fig:LL} decrease for increasing $D$, in line with our expectations.

\begin{figure}
\centering
\includegraphics[width=85mm,height=45mm]{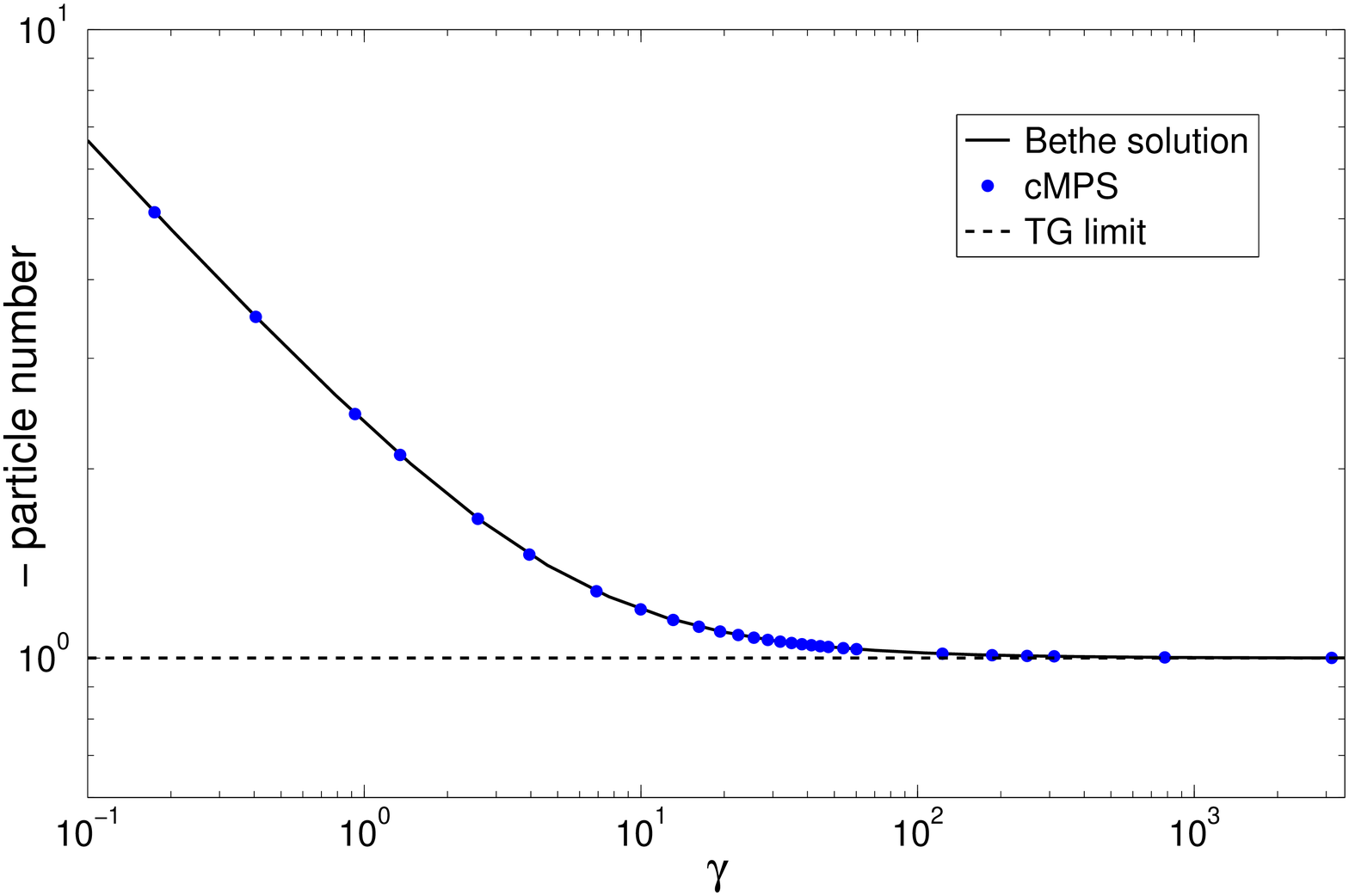}
\caption{\footnotesize{The change in particle number expectation value for the hole state at momentum $p=0$ as a function of $\gamma$, as predicted by the Bethe solution \cite{2012arXiv1210.8337A} and with the cMPS excitation ansatz.}}
\label{fig:LL3}
\end{figure}

Our ansatz does not depend on $\gamma$ and works in principle equally well for both limiting cases $\gamma\rightarrow\infty$ and $\gamma\rightarrow0$, as is confirmed by the results for small values of $\gamma$ in the Supplementary Material. However, for small $\gamma$ ($\gamma\lesssim3$) we observe a transition of the topological hole excitation to a solitonic excitation.  
This observation was first discussed in Refs.~\onlinecite{1976TMP....28..615K,Ishikawa} or recently received renewed attention  following the experimental realization \citep{2012arXiv1204.3960S,2012arXiv1210.8337A,PhysRevLett.101.120406}. In the Tonks-Girardeau limit (large $\gamma$) \citep{1960JMP.....1..516G}, the hole excitation maps exactly to the fermion-like hole excitation and produces a change in particle number of $-1$. For very small $\gamma$, the hole excitation is related to the classical dark soliton of the non-linear Schr\"{o}dinger equation \citep{1971JLTP....4..441T} and produces a large change in particle number \citep{2012arXiv1210.8337A}. The relative change in particle number can easily be calculated using our ansatz. It is illustrated in  Fig.~\ref{fig:LL3} for momentum $p=0$, and the results perfectly coincide with the recent exact calculations \citep{2012arXiv1210.8337A}.

\begin{figure}
\centering
\includegraphics[width=85mm,height=45mm]{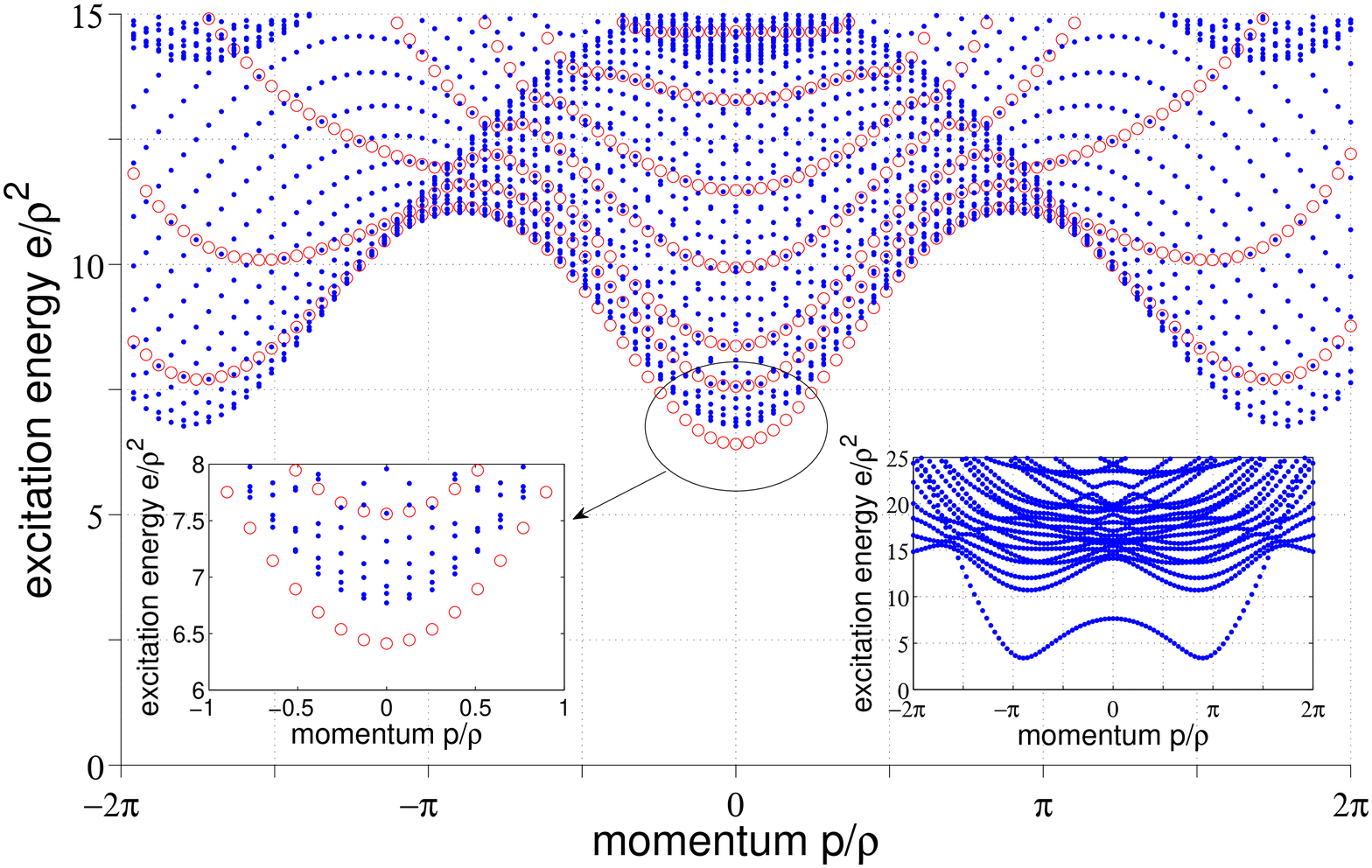}
\includegraphics[width=85mm,height=45mm]{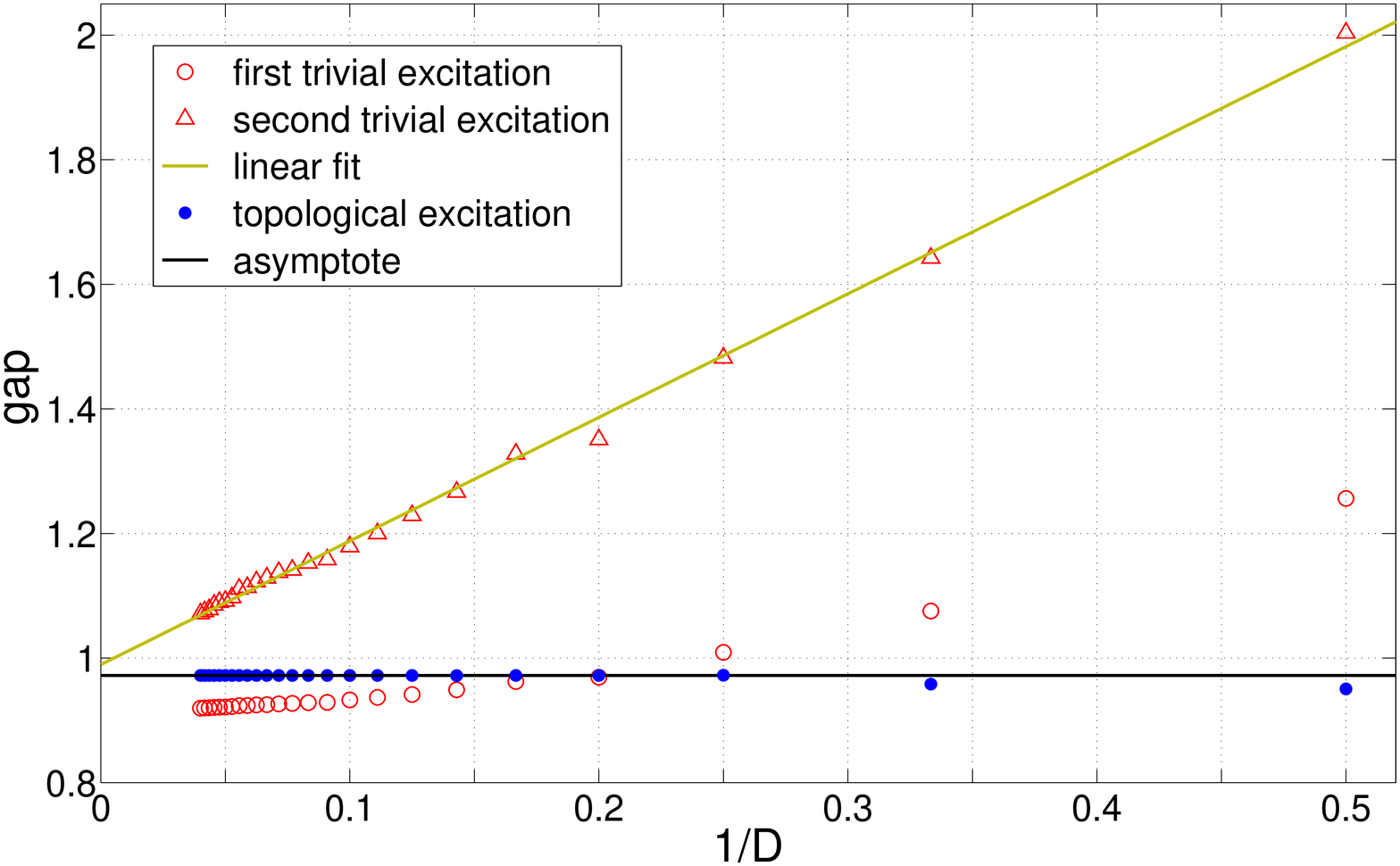}
\caption{\footnotesize{Top: Dispersion relation of $\hat{H}'$ with $D=22$ for $\gamma \approx 26.4$, $\mu = 1$ and $u = 1$  . Blue dots are the pairwise sum of all topological excitations. Red circles are trivial excitations and the right insert shows the topological spectrum. Bottom: Convergence of the two lowest trivial eigenvalues and the lowest two-kink excitation energy as a function of $D^{-1}$ at $p = 0$.}}
\label{fig:pairing}
\end{figure}

Next, we study a non-integrable model obtained by adding a pairing term to $\hat{H}_{\rm{LL}}$ that breaks the $\mathsf{U}(1)$ symmetry down to a residual $\mathds{Z}_{2}$ symmetry
\begin{equation}
\hat{H}' = \hat{H}_{\rm{LL}} + \int_{-\infty}^{\infty} dx \;\left(u \hat{\psi}^{\dagger}\hat{\psi}^{\dagger} + \bar{u} \hat{\psi}\hat{\psi} \right) \nonumber.
\end{equation}
As long as $\mu>0$ the cMPS ground state approximation $\ket{\Psi(Q,R)}$ spontaneously breaks the $\mathds{Z}_{2}$ symmetry for any nonzero pairing strength $u\in\mathds{C}_0$, and a second ground state $\ket{\Psi(Q,-R)}$ is obtained. This also opens up a gap in the spectrum. The magnitude of the order parameter $\braket{\Psi(\bar{Q},\bar{R})|\hat{\psi}(x)|\Psi(Q,R)}$ is determined by the competing effects of the pairing term and the repulsive interaction. As an example we consider an intermediate parameter range where we have found strong evidence that the lowest lying excitation in the trivial spectrum is a bound state. In the top plot of Fig.~\ref{fig:pairing}, the lowest lying trivial excitations of $\hat{H}'$ for $\gamma \approx 26.4$, $\mu = 1$ and $u = 1$ with $D = 22$ are shown with red circles.  Due to the symmetry breaking, we can again construct topological excitations (right insert of Fig.~\ref{fig:pairing}), for which the lowest lying eigenvalues constitute an isolated branch that can be interpreted as the kink excitation that interpolates between the two degenerate ground states. The blue dots in the trivial spectrum of Fig.~\ref{fig:pairing} are obtained by considering all possible pairs of two such topological excitations by adding their momentum and energy. Around momentum zero, the lowest lying trivial excitation produced by the trivial ansatz lies well below the two-kink continuum starting at twice the kink mass.

It is tempting to interpret this excitation as a bound state of two kinks, and this claim is further supported by considering the convergence behavior of (twice) the kink mass and the two lowest trivial excitation energies as a function of $D$ in the bottom plot of Fig.~\ref{fig:pairing}. The kink is the elementary excitation; it has a negative error that vanishes quickly for increasing $D$. As in the LL case, unbound multi-particle states are not expected to converge quickly, because the different constituents are confined within a spatial region, the size of which is determined by the bond dimension
\footnote{In fact, Fig.~\ref {fig:pairing} strongly suggests that the convergence rate is $D^{-1}$}. For the energy of a two-particle state, we thus expect a positive finite size error. The energy of a bound state should also exhibit a large positive error as long as the spatial support of the excitation ansatz is below the binding length, but should then stagnate when the binding length is exceeded by increasing $D$.  Fig.~\ref{fig:pairing} confirms that the lowest lying energy of the trivial ansatz converges to a value well below twice the kink mass in a way that is characteristic for a bound state.

In conclusion, we have introduced a variational ansatz for the study of elementary excitations for one-dimensional quantum field theories, based on continuous matrix product states. The ansatz produces very accurate results even for the critical Lieb-Liniger model. The local aspect of the ansatz made the solitonic feature of the lowest lying hole eigenstates of the LL model very explicit, and the particle number in those solitons could be reproduced to great accuracy. We also applied the method to a gapped non-integrable variant of LL, and obtained strong evidence for the existence of a bound state. As an outlook, we can remark that this method provides a stepping stone to a further understanding of the low-energy dynamics of one-dimensional interacting quantum field theories. In order to calculate \textit{e.g.} dynamical correlation functions, it would be interesting to study scattering states of topological excitations \footnote{In preparation}.

\subsubsection{ACKNOWLEDGMENTS}
\begin{acknowledgments}
We thank G.E.~Astrakharchik and L.P.~Pitaevskii for sharing the particle number data from Ref.~\onlinecite{2012arXiv1210.8337A} and J.S.~Caux, F.~Essler, F.D.M.~Haldane and V.~Korepin for stimulating discussions. We acknowledge support from the EU grants QUERG, by the Austrian FWF SFB grants FoQus and ViCoM, and an Odysseus grant from the Research Foundation Flanders. LV is supported by a Doctoral Scholarship of the Special Research Fund UGent. TJO is supported by the EU grant QFTCMPS and the the cluster of excellence EXC 201 Quantum Engineering and Space-Time Research.
\end{acknowledgments}

\bibliographystyle{achemso}

\bibliography{dispersionbib}




\clearpage

\onecolumngrid
\appendix*

\section{Supplementary Material for\\ "Particles, Holes and Solitons: A Matrix Product State Approach"}
\begin{small}Damian Draxler$^{1}$, Jutho Haegeman$^{2}$, Tobias J.~Osborne$^{3}$, Vid Stojevic$^{12}$, Laurens Vanderstraeten${^2}$, Frank Verstraete$^{12}$
\begin{center}
\textit{$^1$ Vienna Center for Quantum Science, Universit\"at Wien, Boltzmanngasse 5, A-1090 Wien, Austria\\
$^2$ Ghent University, Department of Physics and Astronomy, Krijgslaan 281- S9, B-9000 Ghent, Belgium\\
$^3$ Leibniz Universit\"{a}t Hannover, Institute of Theoretical Physics and Riemann center for geometry and physics, Appelstrasse 2, D-30167 Hannover, Germany}
\end{center}
\end{small}

\maketitle
In this Supplementary Material we explain in full detail how to calculate expectation values of bosonic field operators with the variational states introduced in this paper. In particular, we will derive the effective Hamiltonian for the Lieb-Liniger model $\hat{H}_{\mathrm{LL}}$ and for the non-integrable Hamiltonian $\hat{H}'$ that was introduced in the paper. Moreover, some more results for the spectrum of $\hat{H}_{\mathrm{LL}}$ for small values of the interaction strength are shown. The variational ansatz states for excitations in a uniform system are defined as 
\begin{equation}
 |\Phi_{p}(V,W)\rangle := \int_{-\infty}^{\infty} \mathrm{d}x \,\mathrm{e}^{\mathrm{i}px} v_{L}^{\dagger}\hat{U}_{1}(-\infty,x)\Big( V\otimes\mathds{1} + W\otimes\hat{\psi}^{\dagger}(x) \Big) \hat{U}_{2}(x,\infty)v_{R}|\Omega\rangle \;, \label{eq:Phi}
\end{equation}
with  $V,W\in\mathds{C}^{D\times D}$ and $\hat{U}_{a}(x,y)=\mathcal{P}\exp(\int_{x}^{y}\mathrm{d}z\,(Q_{a}\otimes\mathds{1}+R_{a}\otimes\hat{\psi}^{\dagger}(z)))$ for $a=1,2$ where $(Q_{a},R_{a})$ are two equally good ground states. The field operators satisfy the usual commutation relaltion $[\hat{\psi}(x),\hat{\psi}^{\dagger}(y)] = \delta (x-y)$.  We assume that all eigenvalues of the ground state transfer matrix $T_{aa} = Q_{a}\otimes\mathds{1}+\mathds{1}\otimes\bar{Q}_{a}+R_{a}\otimes\bar{R}_{a}$ with $a=1,2$ have  non-positive real part and there is a non-degenerate zero eigenvalue with corresponding left and right zero eigenvectors, $\langle l_{a}|T_{aa}=0$, $T_{aa}|r_{a}\rangle=0$. Moreover, it is possible to work in a gauge such that $\langle l_{a}|$ always reshapes to $\mathds{1}_{D\times D}$ and $|r_{a}\rangle$ to the right $(D\times D)$-dimensional density matrix $r_{a}$. The boundary vectors are chosen such that the ground state is normalised, i.e.  $\langle\Psi(\bar{Q}_{a},\bar{R}_{a})|\Psi(Q_{a},R_{a})\rangle = (v_{L}^{\dagger}\otimes v_{L}^{\top})|r_{a}\rangle\langle l_{a}|\left(v_{R}\otimes\bar{v}_{R}\right) = 1$.  When $\ket{\Psi(Q_1,R_1)}$ and $\ket{\Psi(Q_2,R_2)}$ represent inequivalent ground states, the mixed transfer matrix $T_{12}$ has only eigenvalues with a strictly negative real part. Vice versa, if the mixed transfer matrix has an eigenvalue with real part zero, then there exists a gauge transform $G$ such that $Q_1=G^{-1} Q_2G$ and $R_1=G^{-1} R_2G$. We will always assume that $(Q_1,R_1)$ and $(Q_2,R_2)$ either parametrize different ground states, or that they are identical such that $G=1$ and $T_{11}=T_{12}=T_{22}$. All of these statements are analogous to well-known results for the case of matrix product states [see \textit{e.g.}\ D.~Perez-Garcia, M.M.~Wolf, M.~Sanz, F.~Verstraete, J.I.~Cirac, Phys.\ Rev.\ Lett.\ \textbf{100}, 167202 (2008)].

To calculate expectation values, it is useful to know how an annihilation operator acts on the states of Eq.~\eqref{eq:Phi}, 
\begin{align}
\hat{\psi}(y)  |\Phi_{p}(V,W)\rangle = &\int_{y}^{\infty} \mathrm{d}x \,\mathrm{e}^{\mathrm{i}px} v_{L}^{\dagger}\hat{U}_{1}(-\infty,y)R_{1}\hat{U}_{1}(y,x)\Big( V\otimes\mathds{1} + W\otimes\hat{\psi}^{\dagger}(x) \Big) \hat{U}_{2}(x,\infty)v_{R}|\Omega\rangle \nonumber \\
& + \int_{-\infty}^{y} \mathrm{d}x \,\mathrm{e}^{\mathrm{i}px} v_{L}^{\dagger}\hat{U}_{1}(-\infty,x)\Big( V\otimes\mathds{1} + W\otimes\hat{\psi}^{\dagger}(x) \Big)\hat{U}_{2}(x,y)R_{2}\hat{U}_{2}(y,\infty)v_{R}|\Omega\rangle \nonumber \\
& +  \mathrm{e}^{\mathrm{i}py}v_{L}^{\dagger}\hat{U}_{1}(-\infty,y)W\hat{U}_{2}(y,\infty)v_{R}|\Omega\rangle \:.\label{eq:psiPhi}
\end{align}
The same can be done for two annihilation operators $\hat{\psi}(z)$, $\hat{\psi}(y)$ where we assume $z\leq y$,
\begin{align}
\hat{\psi}(z) \hat{\psi}(y)  |\Phi_{p}(V,W)\rangle = &\int_{y}^{\infty} \mathrm{d}x \,\mathrm{e}^{\mathrm{i}px} v_{L}^{\dagger}\hat{U}_{1}(-\infty,z)R_{1}\hat{U}_{1}(z,y)R_{1}\hat{U}_{1}(y,x)\Big( V\otimes\mathds{1} + W\otimes\hat{\psi}^{\dagger}(x) \Big) \hat{U}_{2}(x,\infty)v_{R}|\Omega\rangle \nonumber \\
& + \int_{z}^{y} \mathrm{d}x \,\mathrm{e}^{\mathrm{i}px} v_{L}^{\dagger}\hat{U}_{1}(-\infty,z)R_{1}\hat{U}_{1}(z,x)\Big( V\otimes\mathds{1} + W\otimes\hat{\psi}^{\dagger}(x) \Big)\hat{U}_{2}(x,y)R_{2}\hat{U}_{2}(y,\infty)v_{R}|\Omega\rangle \nonumber \\
& +  \int_{-\infty}^{z} \mathrm{d}x \,\mathrm{e}^{\mathrm{i}px} v_{L}^{\dagger}\hat{U}_{1}(-\infty,x)\Big( V\otimes\mathds{1} + W\otimes\hat{\psi}^{\dagger}(x) \Big)\hat{U}_{2}(x,z)R_{2}\hat{U}_{2}(z,y)R_{2}\hat{U}_{2}(y,\infty)v_{R}|\Omega\rangle \nonumber \\
& +  \mathrm{e}^{\mathrm{i}pz}v_{L}^{\dagger}\hat{U}_{1}(-\infty,z)W\hat{U}_{2}(z,y)R_{2}\hat{U}_{2}(y,\infty)v_{R}|\Omega\rangle \nonumber \\
& +  \mathrm{e}^{\mathrm{i}py}v_{L}^{\dagger}\hat{U}_{1}(-\infty,z)R_{1}\hat{U}_{1}(z,y)W\hat{U}_{2}(y,\infty)v_{R}|\Omega\rangle \:.\label{eq:psipsiPhi}
\end{align}
For the kinetic energy we need to know how $\frac{d\hat{\psi}(y)}{dy}$ acts on Eq.~\eqref{eq:Phi}, i.e. we have to take the derivative of Eq.~\eqref{eq:psiPhi}, where we use $\mathrm{d} \hat{U}_a(x,y)/\mathrm{d} y=\hat{U}_{a}(x,y)[Q_a \otimes \hat{\openone} + R_a \otimes \hat{\psi}^{\dagger}(y)]$ and $\mathrm{d} \hat{U}_a(y,x)/\mathrm{d} y=-[Q_a \otimes \hat{\openone} + R_a \otimes \hat{\psi}^{\dagger}(y)]\hat{U}_{a}(y,x)$ in order to obtain
\begin{align}
\frac{d\hat{\psi}(y)}{dy}  |\Phi_{p}(V,W)\rangle = &\int_{y}^{\infty} \mathrm{d}x \,\mathrm{e}^{\mathrm{i}px} v_{L}^{\dagger}\hat{U}_{1}(-\infty,y)\left[Q_{1},R_{1}\otimes\hat{\openone}\right]\hat{U}_{1}(y,x)\Big( V\otimes\mathds{1} + W\otimes\hat{\psi}^{\dagger}(x) \Big) \hat{U}_{2}(x,\infty)v_{R}|\Omega\rangle \nonumber \\
& + \int_{-\infty}^{y} \mathrm{d}x \,\mathrm{e}^{\mathrm{i}px} v_{L}^{\dagger}\hat{U}_{1}(-\infty,x)\Big( V\otimes\mathds{1} + W\otimes\hat{\psi}^{\dagger}(x) \Big)\hat{U}_{2}(x,y)\left[Q_{2},R_{2}\right]\hat{U}_{2}(y,\infty)v_{R}|\Omega\rangle \nonumber \\
& +  v_{L}^{\dagger}\hat{U}_{1}(-\infty,y)\Big( \left(VR_{2} - R_{1}V\right) + (Q_1 W-W Q_2) + \mathrm{i}pW\Big)\hat{U}_{2}(y,\infty)v_{R}|\Omega\rangle \:.\label{eq:dpsiPhi}
\end{align}
One can check that a number of potentially problematic (infinite norm) terms which have a creation operator $\hat{\psi}^{\dagger}(y)$ at the fixed position $y$ all nicely cancel.

Expectation values can now easily be calculated by taking the overlap of the above expressions.
We start with the scalar product of Eq.~\eqref{eq:Phi}
\begin{align}
\langle\Phi_{p}(\overline{V},\overline{W})|\Phi_{p^{\prime}}(V,W)\rangle = &\int_{-\infty}^{\infty}  \mathrm{d}x \,\mathrm{e}^{\mathrm{i}p^{\prime}x} \int_{x}^{\infty} \mathrm{d}y \,\mathrm{e}^{-\mathrm{i}py} \langle l_{1}|\Big( V\otimes\mathds{1} + W\otimes\overline{R}_{1} \Big)\mathrm{e}^{(y-x)T_{12}}\Big( \mathds{1}\otimes\overline{V} + R_{2}\otimes\overline{W} \Big)|r_{2}\rangle \nonumber \\
&+ \int_{-\infty}^{\infty}  \mathrm{d}x \,\mathrm{e}^{\mathrm{i}p^{\prime}x} \int_{-\infty}^{x} \mathrm{d}y \,\mathrm{e}^{-\mathrm{i}py} \langle l_{1}|\Big( \mathds{1}\otimes\overline{V} + R_{1}\otimes\overline{W} \Big)\mathrm{e}^{(x-y)T_{21}}\Big( V\otimes\mathds{1} + W\otimes\overline{R}_{2} \Big)|r_{2}\rangle \nonumber \\
&+ \int_{-\infty}^{\infty}  \mathrm{d}x \,\mathrm{e}^{\mathrm{i}(p^{\prime}-p)x} \langle l_{1}|W\otimes\overline{W}|r_{2}\rangle\nonumber\\
=&2\pi\delta(p'-p) \Bigg[\int_{0}^{\infty} \mathrm{d}y \,\langle l_{1}|\Big( V\otimes\mathds{1} + W\otimes\overline{R}_{1} \Big)\mathrm{e}^{y(T_{12}-\mathrm{i}p)}\Big( \mathds{1}\otimes\overline{V} + R_{2}\otimes\overline{W} \Big)|r_{2}\rangle \nonumber \\
&\qquad\qquad\qquad+ \int_{0}^{+\infty} \mathrm{d}y \, \langle l_{1}|\Big( \mathds{1}\otimes\overline{V} + R_{1}\otimes\overline{W} \Big)\mathrm{e}^{y(T_{21}+\mathrm{i}p)}\Big( V\otimes\mathds{1} + W\otimes\overline{R}_{2} \Big)|r_{2}\rangle \nonumber \\
&\qquad\qquad\qquad+\langle l_{1}|W\otimes\overline{W}|r_{2}\rangle\Bigg] \; .\label{eq:scalarPhi}
\end{align}
The non-local character of the perturbation $(V,W)$ to the ground state results in two non-local terms in the above expression, connected by the mixed transfer matrix $T_{12}$. Throughout this supplementary material, we will have to compute many integrals of the form $\int_{0}^{+\infty} \mathrm{e}^{x(T_{ab}\pm \mathrm{i}p)}\,\mathrm{d}x$. For the mixed transfer matrix in the expression above, in case of two inequivalent ground states, this integral is well defined and results in $-(T_{ab}\pm \mathrm{i}p)^{-1}$, since all eigenvalues have a strictly negative real part. We then obtain
\begin{align}
\langle\Phi_{p}(\overline{V},\overline{W})|\Phi_{p^{\prime}}(V,W)\rangle = &2\pi\delta(p'-p) \Bigg[\langle l_{1}|\Big( V\otimes\mathds{1} + W\otimes\overline{R}_{1} \Big)(-T_{12}+\mathrm{i}p)^{-1}\Big( \mathds{1}\otimes\overline{V} + R_{2}\otimes\overline{W} \Big)|r_{2}\rangle \nonumber \\
&\qquad\qquad\qquad+ \langle l_{1}|\Big( \mathds{1}\otimes\overline{V} + R_{1}\otimes\overline{W} \Big)(-T_{21}-\mathrm{i}p)^{-1}\Big( V\otimes\mathds{1} + W\otimes\overline{R}_{2} \Big)|r_{2}\rangle \nonumber \\
&\qquad\qquad\qquad+\langle l_{1}|W\otimes\overline{W}|r_{2}\rangle\Bigg] \; .\label{eq:scalarPhi1}
\end{align}
For topologically trivial states [$(Q_1,R_1)=(Q_2,R_2)$], the situation is more complicated. In this case, we can define a unique transfer matrix $T:=T_{12}=T_{11}=T_{22}$ which has a unique eigenvalue zero with corresponding left eigenvector $\bra{l}:=\bra{l_1}=\bra{l_2}$ and right eigenvector $\ket{r}:=\ket{r_1}=\ket{r_2}$. We can then define a projector $P=\openone-\ket{r}\bra{l}$ and compute $\int_{0}^{+\infty} P\mathrm{e}^{x(T\pm \mathrm{i}p)}P\,\mathrm{d}x=-P(T\pm \mathrm{i}p)^{-1}P=-(T\pm \mathrm{i}p)^{\mathsf{P}}$ where we have introduced the notation of superscript $\mathsf{P}$ to denote a kind of pseudo-inverse where the eigenspace zero of the transfer matrix has been projected out. The remaining `singular' part evaluates to $\int_{0}^{+\infty} \ket{r}\bra{l}\mathrm{e}^{x(T\pm \mathrm{i}p)}\ket{r}\bra{l}\,\mathrm{d}x=\ket{r}\bra{l} \int_{0}^{+\infty} \mathrm{e}^{\pm \mathrm{i}px}\,\mathrm{d}x$, and by combining the singular parts of the first and second line in Eq.~\eqref{eq:scalarPhi} we obtain
\begin{align}
\langle\Phi_{p}(\overline{V},\overline{W})|\Phi_{p^{\prime}}(V,W)\rangle = &2\pi\delta(p'-p) \Bigg[\langle l|\Big( V\otimes\mathds{1} + W\otimes\overline{R} \Big)(-T+\mathrm{i}p)^{\mathsf{P}}\Big( \mathds{1}\otimes\overline{V} + R\otimes\overline{W} \Big)|r\rangle \nonumber \\
&\qquad\qquad\qquad+ \langle l|\Big( \mathds{1}\otimes\overline{V} + R\otimes\overline{W} \Big)(-T-\mathrm{i}p)^{\mathsf{P}}\Big( V\otimes\mathds{1} + W\otimes\overline{R} \Big)|r\rangle \nonumber \\
&\qquad\qquad\qquad+\langle l|W\otimes\overline{W}|r\rangle+2\pi\delta(p)\langle l|\mathds{1}\otimes\overline{V} + R\otimes\overline{W}\ket{r}\bra{l} V\otimes\mathds{1} + W\otimes\overline{R} |r\rangle\Bigg] \; .\label{eq:scalarPhi1}
\end{align}
The singular term for momentum $p=0$ inside the square brackets can be traced back to the non-orthogonality of the $\ket{\Phi_p(V,W)}$ with the ground state $\ket{\Psi(Q,R)}$, which is given by $\braket{\Psi(Q,R)|\Phi_p(V,W)}=2\pi\delta(p)\langle l|\mathds{1}\otimes\overline{V} + R\otimes\overline{W}\ket{r}$. For topologically trivial excitations at momentum zero, we henceforth restrict to $V$ and $W$ which satisfy $\langle l|\mathds{1}\otimes\overline{V} + R\otimes\overline{W}\ket{r}=0$, so that this singular term vanishes. 

The gauge freedom in the cMPS parametrization induces a redundancy in the parametrization of the states $\ket{\Phi_{p}(V,W)}$, \textit{i.e.} these states are invariant under the additive gauge transformation $V \leftarrow V + Q_{1}X - XQ_{2} + \mathrm{i}pX$ and $W \leftarrow W + R_{1}X - XR_{2}$ for $X\in \mathfrak{gl}(D)=\mathbb{C}^{D\times D}$. This gauge freedom can be used to choose a parametrization that allows us omit the non-local terms in the expressions above, \textit{e.g.} by restriction to representations $(V,W)$ that satisfy
\begin{equation}
\langle l_{1}|\Big( V\otimes\mathds{1} + W\otimes\overline{R}_{1} \Big) = 0.\label{eq:leftgauge}
\end{equation}
This condition is henceforth referred to as the \textit{left gauge condition}. Similarly one can choose instead a right gauge condition by imposing   $\Big( V\otimes\mathds{1} + W\otimes\overline{R}_{2} \Big)|r_{2}\rangle = 0$. With this, Eq.~\eqref{eq:scalarPhi} reduces to the local expression
\begin{equation}
\langle\Phi_{p}(\overline{V},\overline{W})|\Phi_{p^{\prime}}(V,W)\rangle =  2\pi\delta(p-p^{\prime}) \langle l_{1}|W\otimes\overline{W}|r_{2}\rangle \;
\end{equation}
for the topologically trivial and non-trivial cases alike. Note however the subtle difference between both cases. In the topologically non-trivial case (inequivalent ground states $\ket{\Psi(Q_1,R_1)}$ and $\ket{\Psi(Q_2,R_2)}$), it is really all $D^2$ gauge degrees of freedom in $X$ that are used to impose $D^2$ equations that are encoded by the left or right gauge condition. For the topologically trivial case, this is still true for momenta $p\neq 0$. However, at momentum zero the choice $X=\openone$ does not induce any transformation at all, and the gauge transformations can only be used to fulfil $D^2-1$ conditions. The remaining requirement is $\langle l|\mathds{1}\otimes\overline{V} + R\otimes\overline{W}\ket{r}=0$, which expresses orthogonality to the ground state and cannot be imposed by a mere gauge transform. Hence, the left or right gauge condition automatically include the restriction of orthogonality to the ground state for topologically trivial excitations with momentum $p=0$.

Next we will calculate the expectation value of the terms in the Hamiltonian. Evaluating these expressions will involve more integrals of the form $\int_{0}^{+\infty} \mathrm{e}^{x(T_{ab}\pm \mathrm{i}p)}\,\mathrm{d}x$ or $\int_{0}^{+\infty} \mathrm{e}^{xT_{aa}}\,\mathrm{d}x$, and we now introduce the general notation $(T_{ab}\pm \mathrm{i}p)^{\mathsf{P}}$, which is defined as the inverse of $T_{ab}\pm \mathrm{i}p$ in the complement of the zero eigenspace of $T_{ab}$ (which is the whole space $\mathbb{C}^{D^2}$ if $T_{ab}$ doesn't have a zero eigenvalue). When acting with the left or right zero eigenvector of $T_{ab}$ on $(T_{ab}\pm \mathrm{i}p)^{\mathsf{P}}$, it reduces to zero. We will always be able to cancel all singular contributions, either by using the fact that we choose $\langle l|\mathds{1}\otimes\overline{V} + R\otimes\overline{W}\ket{r}=0$ in the case of topologically trivial excitations at momentum zero or by subtracting the ground state energy. Let us elaborate on the latter. There will be contributions where the local operators in the Hamiltonian do not act directly on the perturbation $V$ and $W$. However, the nonlocal effect of this perturbation is expressed by the fact that there is a transfer matrix connecting that Hamiltonian term to the perturbation. The regular part of this transfer matrix is actually responsible for the genuine nonlocal effect of the perturbation, whereas the singular part contains the projection $\ket{r}\bra{l}$ and just returns the ground state energy density with a singular volume integral. It is only at the very end, by subtracting the total ground state energy (\textit{i.e.} the energy density times the volume) that these singular contributions disappear. This also requires that $\ket{\Psi(Q_1,R_1)}$ and $\ket{\Psi(Q_2,R_2)}$ have identical energy densities in the case of topologically nontrivial excitations. However, keeping in mind this final step, we will not include these singular contributions in the expressions below. Note that the use of the left gauge condition allows to omit additional non-singular terms.

The expectation value of the density operator is obtained by computing the overlap of Eq.~\eqref{eq:psiPhi} with itself
\begin {align}
\int_{-\infty}^{\infty} \mathrm{d}x& \,\langle\Phi_{p}(\overline{V},\overline{W})| \hat{\psi}^{\dagger}(x)\hat{\psi}(x) |\Phi_{p^{\prime}}(V,W)\rangle = 2\pi\delta(p-p^{\prime}) \times \nonumber \\
&\Big\{\langle l_{1}|\Big[ R_{1}\otimes\overline{R}_{1}\Big(-T_{11}\Big)^{\mathsf{P}}W\otimes\overline{W} + W\otimes\overline{W}\Big(-T_{22}\Big)^{\mathsf{P}}R_{2}\otimes\overline{R}_{2}  \nonumber \\
& \:\:\:\:\:\:\:+ R_{1}\otimes\overline{R}_{1}\Big(-T_{11}\Big)^{\mathsf{P}}\Big( V\otimes\mathds{1} + W\otimes\overline{R}_{1}\Big)\Big(-T_{21}+\mathrm{i}p\Big)^{\mathsf{P}}\Big( \mathds{1}\otimes\overline{V} + R_{2}\otimes\overline{W} \Big)  \nonumber \\
& \:\:\:\:\:\:\:+R_{1}\otimes\overline{R}_{1}\Big(-T_{11}\Big)^{\mathsf{P}}\Big( \mathds{1}\otimes\overline{V} + R_{1}\otimes\overline{W} \Big)\Big(-T_{12}-\mathrm{i}p^{\prime}\Big)^{\mathsf{P}}\Big( V\otimes\mathds{1} + W\otimes\overline{R}_{2}\Big)   \nonumber \\  
& \:\:\:\:\:\:\:+R_{1}\otimes\overline{W}\Big(-T_{12}-\mathrm{i}p^{\prime}\Big)^{\mathsf{P}}\Big( V\otimes\mathds{1} + W\otimes\overline{R}_{2}\Big)    \nonumber \\
& \:\:\:\:\:\:\:+W\otimes\overline{R}_{1}\Big(-T_{21}+\mathrm{i}p\Big)^{\mathsf{P}}\Big( \mathds{1}\otimes\overline{V} + R_{2}\otimes\overline{W} \Big) + W\otimes\overline{W} \Big]|r_{2}\rangle\Big\} \;.\label{eq:scalarpsiPhi}
\end{align} 

The expectation value of the interaction energy of the $\delta$-interaction is calculated similarly by taking the overlap of Eq.~\eqref{eq:psipsiPhi} in the limit $z\rightarrow y$,
\begin{align}
\int_{-\infty}^{\infty} \mathrm{d}x& \,\langle\Phi_{p}(\overline{V},\overline{W})|\hat{\psi}^{\dagger}(x)\hat{\psi}^{\dagger}(x)\hat{\psi}(x)\hat{\psi}(x) |\Phi_{p^{\prime}}(V,W)\rangle = 2\pi\delta(p-p^{\prime}) \times \nonumber \\
&\Big\{\langle l_{1}|\Big[ R^{2}_{1}\otimes\overline{R}^{2}_{1}\Big(-T_{11}\Big)^{\mathsf{P}}W\otimes\overline{W} + W\otimes\overline{W}\Big(-T_{22}\Big)^{\mathsf{P}}R^{2}_{2}\otimes\overline{R}^{2}_{2}  \nonumber \\
& \:\:\:\:\:\:\:+ R^{2}_{1}\otimes\overline{R}^{2}_{1}\Big(-T_{11}\Big)^{\mathsf{P}}\Big( V\otimes\mathds{1} + W\otimes\overline{R}_{1}\Big)\Big(-T_{21}+\mathrm{i}p\Big)^{\mathsf{P}}\Big( \mathds{1}\otimes\overline{V} + R_{2}\otimes\overline{W} \Big)  \nonumber \\
& \:\:\:\:\:\:\:+R^{2}_{1}\otimes\overline{R}^{2}_{1}\Big(-T_{11}\Big)^{\mathsf{P}}\Big( \mathds{1}\otimes\overline{V} + R_{1}\otimes\overline{W} \Big)\Big(-T_{12}-\mathrm{i}p^{\prime}\Big)^{\mathsf{P}}\Big( V\otimes\mathds{1} + W\otimes\overline{R}_{2}\Big)  \nonumber \\  
& \:\:\:\:\:\:\:+R^{2}_{1}\otimes(\overline{R}_{1}\overline{W} + \overline{W} \,\overline{R}_{2}) \Big(-T_{12}-\mathrm{i}p^{\prime}\Big)^{\mathsf{P}}\Big( V\otimes\mathds{1} + W\otimes\overline{R}_{2}\Big)  \nonumber \\ 
& \:\:\:\:\:\:\:+ (R_{1}W + WR_{2})\otimes\overline{R}^{2}_{1}\Big(-T_{21}+\mathrm{i}p\Big)^{\mathsf{P}}\Big( \mathds{1}\otimes\overline{V} + R_{2}\otimes\overline{W} \Big)  \nonumber \\ 
& \:\:\:\:\:\:\:+ (R_{1}W + WR_{2})\otimes(\overline{R}_{1}\overline{W} + \overline{W} \,\overline{R}_{2}) \Big]|r_{2}\rangle \Big\} \;.
\end{align}      
For the kinetic energy we evaluate the overlap of Eq.~\eqref{eq:dpsiPhi},
\begin{align}
\int_{-\infty}^{\infty} \mathrm{d}x& \,\langle\Phi_{p}(\overline{V},\overline{W})|\frac{d\hat{\psi}^{\dagger}(x)}{dx}\frac{d\hat{\psi}(x)}{dx} |\Phi_{p^{\prime}}(V,W)\rangle = 2\pi\delta(p-p^{\prime}) \times \nonumber \\
&\Big\{\langle l_{1}|\Big[ [Q_1,R_{1}]\otimes[\overline{Q}_1,\overline{R}_{1}]\Big(-T_{11}\Big)^{\mathsf{P}}W\otimes\overline{W} + W\otimes\overline{W}\Big(-T_{22}\Big)^{\mathsf{P}}[Q_2,R_{2}]\otimes[\overline{Q}_2,\overline{R}_{2}]  \nonumber \\
& \:\:\:\:\:\:\:+ [Q_1,R_{1}]\otimes[\overline{Q}_1,\overline{R}_{1}]\Big(-T_{11}\Big)^{\mathsf{P}}\Big( V\otimes\mathds{1} + W\otimes\overline{R}_{1}\Big)\Big(-T_{21}+\mathrm{i}p\Big)^{\mathsf{P}}\Big( \mathds{1}\otimes\overline{V} + R_{2}\otimes\overline{W} \Big)  \nonumber \\
& \:\:\:\:\:\:\:+[Q_1,R_{1}]\otimes[\overline{Q}_1,\overline{R}_{1}]\Big(-T_{11}\Big)^{\mathsf{P}}\Big( \mathds{1}\otimes\overline{V} + R_{1}\otimes\overline{W} \Big)\Big(-T_{12}-\mathrm{i}p^{\prime}\Big)^{\mathsf{P}}\Big( V\otimes\mathds{1} + W\otimes\overline{R}_{2}\Big)  \nonumber \\  
& \:\:\:\:\:\:\:+((Q_1 W-W Q_2)+(VR_{2}-R_{1}V)+\mathrm{i}pW)\otimes[\overline{Q}_1,\overline{R}_{1}]\Big(-T_{21}+\mathrm{i}p\Big)^{\mathsf{P}} \Big( \mathds{1}\otimes\overline{V} + R_{2}\otimes\overline{W} \Big) \nonumber \\
& \:\:\:\:\:\:\:+[Q_1,R_{1}]\otimes((\overline{Q}_1\overline{W}-\overline{W}\overline{Q}_2) + (\overline{V}\,\overline{R}_{2}-\overline{R}_{1}\overline{V})-\mathrm{i}p\overline{W})\Big(-T_{12}-\mathrm{i}p^{\prime}\Big)^{\mathsf{P}}\Big( V\otimes\mathds{1} + W\otimes\overline{R}_{2}\Big) \nonumber \\
& \:\:\:\:\:\:\:+((Q_1W-WQ_2)+(VR_{2}-R_{1}V)+\mathrm{i}p W)\otimes((\overline{Q}_1\overline{W}-\overline{W}\overline{Q}_2) + (\overline{V}\,\overline{R}_{2}-\overline{R}_{1}\overline{V})-\mathrm{i}p \overline{W}) \Big]|r_{2}\rangle \Big\} \;.
\end{align}

Finally we calculate the expectation values of the pairing terms which appear in the non-integrable Hamiltonian introduced in this paper. 
\begin{align}
\int_{-\infty}^{\infty} \mathrm{d}x& \,\langle\Phi_{p}(\overline{V},\overline{W})|\hat{\psi}^{\dagger}(x)\hat{\psi}^{\dagger}(x) |\Phi_{p^{\prime}}(V,W)\rangle = 2\pi\delta(p-p^{\prime}) \times \nonumber \\
&\Big\{\langle l_{1}|\Big[ \mathds{1}\otimes\overline{R}^{2}_{1}\Big(-T_{11}\Big)^{\mathsf{P}}W\otimes\overline{W} + W\otimes\overline{W}\Big(-T_{22}\Big)^{\mathsf{P}}\mathds{1}\otimes\overline{R}^{2}_{2}  \nonumber \\
& \:\:\:\:\:\:\:+ \mathds{1}\otimes\overline{R}^{2}_{1}\Big(-T_{11}\Big)^{\mathsf{P}}\Big( V\otimes\mathds{1} + W\otimes\overline{R}_{1}\Big)\Big(-T_{21}+\mathrm{i}p\Big)^{\mathsf{P}}\Big( \mathds{1}\otimes\overline{V} + R_{2}\otimes\overline{W} \Big)  \nonumber \\
& \:\:\:\:\:\:\:+\mathds{1}\otimes\overline{R}^{2}_{1}\Big(-T_{11}\Big)^{\mathsf{P}}\Big( \mathds{1}\otimes\overline{V} + R_{1}\otimes\overline{W} \Big)\Big(-T_{12}-\mathrm{i}p^{\prime}\Big)^{\mathsf{P}}\Big( V\otimes\mathds{1} + W\otimes\overline{R}_{2}\Big)  \nonumber \\  
& \:\:\:\:\:\:\:+\mathds{1}\otimes(\overline{R}_{1}\overline{W} + \overline{W} \,\overline{R}_{2}) \Big(-T_{12}-\mathrm{i}p^{\prime}\Big)^{\mathsf{P}}\Big( V\otimes\mathds{1} + W\otimes\overline{R}_{2}\Big)\Big]|r_{2}\rangle \Big\} \;
\end{align}
and
\begin{align}
\int_{-\infty}^{\infty} \mathrm{d}x& \,\langle\Phi_{p}(\overline{V},\overline{W})|\hat{\psi}(x)\hat{\psi}(x) |\Phi_{p^{\prime}}(V,W)\rangle = 2\pi\delta(p-p^{\prime}) \times \nonumber \\
&\Big\{\langle l_{1}|\Big[ R^{2}_{1}\otimes\mathds{1}\Big(-T_{11}\Big)^{\mathsf{P}}W\otimes\overline{W} + W\otimes\overline{W}\Big(-T_{22}\Big)^{\mathsf{P}}R^{2}_{2}\otimes\mathds{1}  \nonumber \\
& \:\:\:\:\:\:\:+ R^{2}_{1}\otimes\mathds{1}\Big(-T_{11}\Big)^{\mathsf{P}}\Big( V\otimes\mathds{1} + W\otimes\overline{R}_{1}\Big)\Big(-T_{21}+\mathrm{i}p\Big)^{\mathsf{P}}\Big( \mathds{1}\otimes\overline{V} + R_{2}\otimes\overline{W} \Big)  \nonumber \\
& \:\:\:\:\:\:\:+R^{2}_{1}\otimes\mathds{1}\Big(-T_{11}\Big)^{\mathsf{P}}\Big( \mathds{1}\otimes\overline{V} + R_{1}\otimes\overline{W} \Big)\Big(-T_{12}-\mathrm{i}p^{\prime}\Big)^{\mathsf{P}}\Big( V\otimes\mathds{1} + W\otimes\overline{R}_{2}\Big)  \nonumber \\  
& \:\:\:\:\:\:\:+R^{2}_{1}\otimes(\overline{R}_{1}\overline{W} + \overline{W} \,\overline{R}_{2}) \Big(-T_{12}-\mathrm{i}p^{\prime}\Big)^{\mathsf{P}}\Big( V\otimes\mathds{1} + W\otimes\overline{R}_{2}\Big)\Big]|r_{2}\rangle \Big\} \;
\end{align}      

The above expressions define the effective Hamiltonian for the Lieb-Liniger model and for the non-integrable model. The explicit construction of the effective Hamiltonian is done by making use of the fact that these expressions are linear in $V$ and $W$. This allows us to reshape them into  the form $\langle V^{\dagger} W^{\dagger}| H_{\rm{eff}}| V W\rangle$ and $\langle V^{\dagger} W^{\dagger}| N_{\rm{eff}}| V W\rangle$. Since we impose the left gauge condition Eq.~\eqref{eq:leftgauge} we can parametrize $V$ and $W$ in terms of one single matrix $Y\in \mathds{C}^{D\times D}$. We can choose a parametrization such that the effective norm matrix $N_{\rm{eff}}$ is the identity by setting $W = l_{1}^{-1/2} Y r^{-1/2}_{2}$ and $V=-l_1^{-1} R^{\dagger}l_1^{1/2}Yr^{-1/2}_{2}$, where we can also use a gauge for the cMPS such that $l_1=\openone$. This particular parametrization is useful since it transforms the generalised eigenvalue problem for the effective Hamiltonian into an ordinary one and allows for an efficient iterative implementation with computational cost of order $\mathcal{O}(D^3)$ .

Finally, we briefly discuss our results for the Lieb-Liniger model in the weak coupling limit of small $\gamma$. It is well known that in this limit the mean-field approximation derived by Bogoliubov accurately describes the Type \MakeUppercase{\romannumeral 1} excitations. We therefore expect to find this branch in the trivial spectrum of our ansatz (dots in  Fig.~\ref{fig:LL2})  by looking for the excitation which has largest overlap with the state $\int_{-\infty}^{\infty}\mathrm{d}x\,\mathrm{e}^{\mathrm{i} p x} \hat{\psi}^{\dagger}(x) |\Psi(Q,R)\rangle$. These excitations are shown by blue diamonds in  Fig.~\ref{fig:LL2} .  However, the branch of  Type \MakeUppercase{\romannumeral 2} excitations is always found by combining two excitations of the topological spectrum (red stars in  Fig.~\ref{fig:LL2}), irrespective of the interaction strength. The left plot of Fig.~\ref{fig:LL2}  clearly shows that for weak interaction the lowest branch of trivial excitations is separated by a large gap from the energy of the lowest topological excitation (red stars).  This can be understood from the fact that the trivial ansatz badly describes unbound multi-particle excitations, as explained in the paper . For strong interaction the particle number of the topological \textit{hole} excitation is $-1$ and therefore also the trivial ansatz gives a fairly good description of the Type \MakeUppercase{\romannumeral 2} branch (Fig. 1. in the paper). However, for small $\gamma$ the  topological \textit{hole} excitation experiences a diverging particle number and the trivial ansatz is no longer able to accurately describe this excitation branch which leads to the large energy gap in Fig.~\ref{fig:LL2}.

\begin{figure}
\includegraphics[width=85mm,height=44mm]{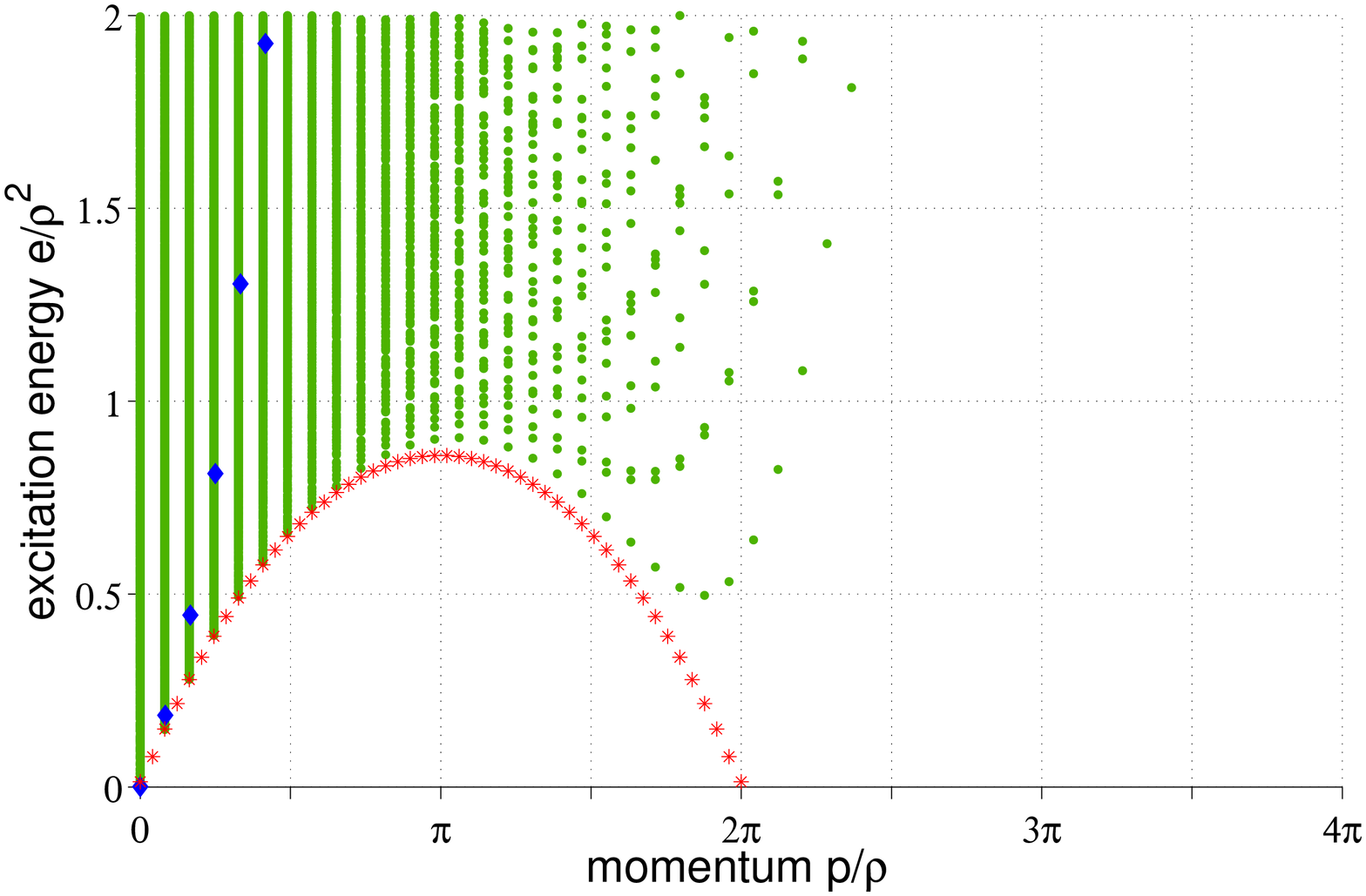}\includegraphics[width=85mm,height=44mm]{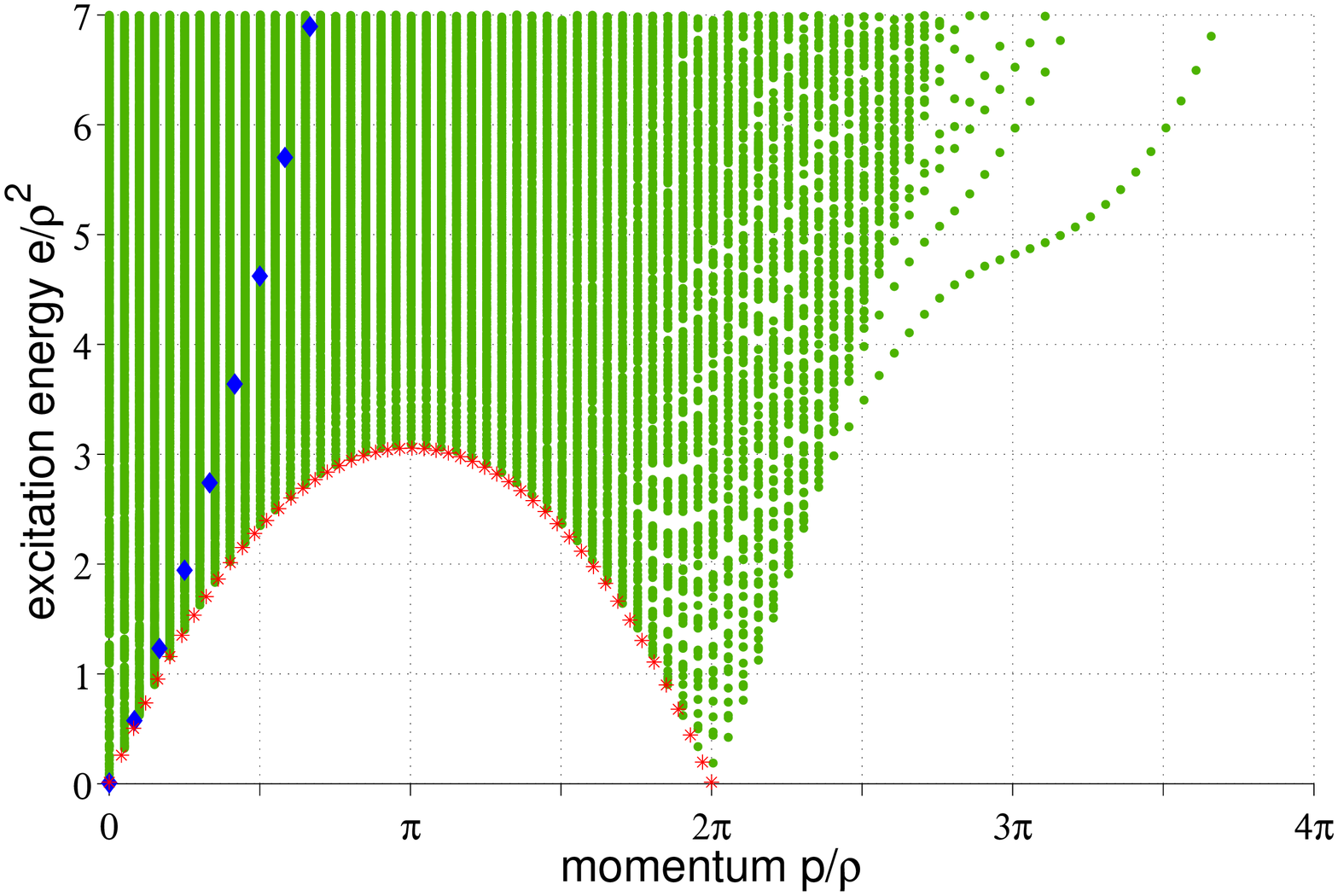}
\caption{\footnotesize{Dispersion relation for the Lieb-Liniger model with $D=64$ for $\gamma =c/\rho\approx  0.1$ and $\gamma\approx 1.35$ (left to right). Bulk excitations (dots) and blue diamonds (Type \MakeUppercase{\romannumeral 1}) are trivial excitations. Red stars (Type \MakeUppercase{\romannumeral 2}) are obtained by combining momenta and energy of two topological  excitations (hole and particle).}}
\label{fig:LL2}
\end{figure}

\end{document}